\documentclass[11pt]{amsart}
\usepackage{amsmath}
\usepackage{amsthm}
\usepackage{amsfonts}
\usepackage{amssymb}
\usepackage{epic}
\usepackage{eepic}

\newtheorem{thm}{Theorem}
\newtheorem{prop}{Proposition}

\theoremstyle{definition}
\newtheorem{defn}{Definition}

\theoremstyle{remark}

\newcommand{\beq}{\begin{equation}}
\newcommand{\eeq}{\end{equation}}

\newcommand{\pa}{\partial}

\newcommand{\ra}{\rightarrow}

\newcommand{\fr}[2]{{\textstyle \frac{#1}{#2} }}

\newcommand{\al}{\alpha}
\newcommand{\be}{\beta}
\newcommand{\ga}{\gamma}
\newcommand{\Ga}{\Gamma}
\newcommand{\de}{\delta}
\newcommand{\De}{\Delta}
\newcommand{\ep}{\epsilon}
\newcommand{\ka}{\kappa}
\newcommand{\la}{\lambda}
\newcommand{\om}{\omega}

\newcommand{\ba}{\bar{a}}
\newcommand{\bb}{\bar{b}}

\newcommand{\CA}{{\mathcal A}}

\newcommand{\CD}{{\mathcal D}}

\newcommand{\CH}{{\mathcal H}}
\newcommand{\CI}{{\mathcal I}}

\newcommand{\CK}{{\mathcal K}}

\newcommand{\CM}{{\mathcal M}}

\newcommand{\CO}{{\mathcal O}}

\newcommand{\CS}{{\mathcal S}}
\newcommand{\CT}{{\mathcal T}}
\newcommand{\CU}{{\mathcal U}}

\newcommand{\FB}{{\mathfrak B}}

\newcommand{\BR}{{\mathbb R}}

\newcommand{\BC}{{\mathbb C}}

\newcommand{\BZ}{{\mathbb Z}}

\newcommand{\gas}{g^{\text{\tiny as}}}

\newcommand{\gin}{g^{\text{\tiny in}}}

\newcommand{\fout}{f^{\text{\tiny out}}}
\newcommand{\gout}{g^{\text{\tiny out}}}

\newcommand{\rf}[1]{(\ref{#1})}

\newcommand{\aufz}
{\begin{list}{$\bullet$}{\topsep0cm \itemsep0cm \parsep0cm}}
\newcommand{\eaufz}{\end{list}}
\newcounter{num}
\newcommand{\remlst}{\begin{list}
{(\arabic{num})}{\usecounter{num}\topsep0cm \itemsep0cm \parsep0cm}}
\begin{document}
\thispagestyle{empty}
\hspace*{\fill} DIAS-STP-99-02\\[.5cm]
\title{The deformed two-dimensional black hole}
\author{\sc J. Teschner}
\address{School for theoretical Physics, Dublin Institute for Advanced
Studies, 10 Burlington Road, Dublin 4, Ireland}
\email{teschner@stp.dias.ie}

\begin{abstract}
A deformation of the wave equation on a two-dimensional black hole 
is considered as a toy-model for possible gravitational or stringy 
nonlocal effects. The deformed wave-equation allows for an initial-value
problem despite being nonlocal. 
The singularity present in the classical geometry is
resolved by the deformation, so that propagation of a wave-packet can be
continued through the classically singular region, ultimately reaching
another asymptotically ``flat'' region.
\end{abstract}

\maketitle

\section{Introduction}

It is an important open question whether a notion of locality can be 
maintained in fundamental theories involving gravity such as e.g. 
string theory. This problem appears to be closely related to the question
whether (pseudo-) Riemannian geometry still makes sense down to arbitrarily 
small distances. 

Arguments have been put forward that indicate the impossibility of 
measuring arbitrarily small distances when combining quantum theory and general
relativity, see e.g. \cite{DFR}\cite{FGR}. If Riemannian geometry can not
be measured then it becomes questionable whether it is a useful concept 
for the description of physical situations in which 
space-time uncertainties are relevant. 

The status of locality in string theory is unclear due to the absence of
a fundamental, background-independent formulation. There are hints indicating
that strings cannot probe arbitrarily small distances due to quantum 
fluctuations of their shape, which are related to the fact that
effective actions for the fields corresponding to the low energy states
of the string are generically\footnote{if $l_s/l_R$ is not 
very small, where $l_s$ is the string length and $l_R$ denotes
the characteristic curvature radius} highly nonlocal. More recent developments
have exhibited on the one hand D-branes as probes that sometimes are
able to resolve smaller distance scales than strings 
\cite{DKPS}\footnote{The situation for $l_s/l_R\sim \CO(1)$ 
it is not clear to the author. It has been proposed in \cite{D} that 
D0-branes behave like {\it point}-particles in some regime with 
$l_s/l_R\sim \CO(1)$ so that the metric probed by these objects would
define the pseudo-Riemannian structure of a background. On the other hand
it appears to the author that the description of D-brane
effective actions in terms of noncommutative geometry that was found in 
some specific backgrounds (reviewed in \cite{D}) should be generic rather
than exceptional.}, 
but have on the other hand discovered new
hints towards fundamental nonlocalities from non-perturbative dualities 
such as the dualities between strings on Anti-de Sitter (AdS) spaces and 
conformal field theories conjectured by
Maldacena (see e.g. \cite{BDHM}). 

Having available more general frameworks for the formulation of field 
theories that take into account fundamental space-time uncertainties or
nonlocal effects may be an important ingredient for the further
development of gravitational or string theories. One rather general framework
that has been proposed in this context is given by the noncommutative
geometry \cite{Co}, where non-commutativity of the operator algebra supposed 
to describe position measurements prevents localization to arbitrarily small 
distances. One thereby naturally obtains space-time uncertainties as 
e.g. in the explicit example proposed in \cite{DFR}. 

There is a generalization of the concept of
geodesic distance for noncommutative
spaces \cite{Co}\cite{FG}, which is based on the observation that the
geodesic distance on  a Riemannian manifold $\CM$ can be reconstructed from the
{\it algebraic} data $(\CA,\CH,\De)$, where $\CA$ is the algebra of (say)
smooth, bounded functions on $\CM$, $\CH$ may be taken as the space 
$L^2(\CM,dv)$ on which $\CA$ acts as multiplication operators and 
$\De$ is the Laplacian on $\CM$ considered as a self-adjoint operator on $\CH$.
Physically this simply means that the geodesic distance can be reconstructed
from the quantum mechanics of point particles on $\CM$. This observation 
leads to a natural definition of geodesic distance for more general
choices of data $(\CA,\CH,\De)$, in which the algebra $\CA$ may be 
noncommutative. A possibility that does not seem to have attracted much
attention is the case in which $\CA$ is still commutative, but 
$\De$ is replaced by some other, maybe nonlocal, self-adjoint operator on 
$\CH$.

Unfortunately,  an
analogous generalization in the case of Minkowskian signature metrics
does not seem to exist presently (to the author's knowledge).
Physically one might guess that reconstructing the Riemannian metric 
from the quantum mechanics on $\CM$ should be replaced by a study of
propagation of wave-packets according to the covariant wave equation.
This suggests that it should ultimately be possible to define 
generalizations of pseudo-Riemannian structures from data such as 
$(\CA,\CH,\square)$,
where now $\square$ is some generalization of the covariant wave operator.
It is in this spirit that deformations of the covariant wave equation 
will be considered as defining a deformed pseudo-Riemannian geometry. 

Alternatively, one may simply take the point of view that nonlocal
deformations of the covariant wave equation 
are one way to parametrize certain modifications
of the propagation of fields on a manifold due to 
gravitational/stringy nonlocal effects. 

The present paper will describe an example of 
a nonlocal deformation of the wave equation in the geometry of a 
two-dimensional black hole. In this example the generalization 
appears exclusively in the nonlocality of the deformed covariant wave operator,
the underlying algebra of functions is {\it commutative}. 
However, the specific deformation studied has its roots in a 
{\it noncommutative} quantum
deformation of the algebra of functions on $SL(2,\BR)$
and was found by generalizing the
relation between $SL(2,\BR)$ and the two-dimensional black hole that exists
classically (e.g. \cite{Wi}\cite{DVV}), as will be further discussed 
elsewhere. By similar constructions one can alternatively obtain 
{\it noncommutative} deformations of
Euclidean/Minkowskian $AdS_3$, the BTZ-black hole, and the
two-dimensional euclidean black hole.
 
The paper starts with a presentation of some results on wave propagation
and scattering in the case of the classical two-dimensional black hole.
Some important results had been obtained in \cite{DVV}, but for the
purpose of comparison with the deformed case it appeared to be necessary
to complete (e.g. by the solution of the Cauchy problem) 
and generalize (to arbitrary mass) the discussion therein. 

The following section then carries out a study of the deformed case along
similar lines as in the classical case. In order to focus on the
physically interesting aspects and to keep the discussion brief, the basic
technical results, being analogous to the classical case, are only announced.
One should note however that their proof requires methods rather different 
from the well-known techniques that can be used in the classical case, so a 
full account of the technicalities will be given elsewhere.

\section{The two-dimensional black hole}

A two-dimensional analogue of the black hole
\footnote{See e.g. \cite{Wi}\cite{DVV} for more extensive discussions} 
can be found as solution of the
equations of motion for the dilaton gravity theory defined by the action
\begin{equation}
S=\int d^2x \sqrt{G} e^{\Phi}\Bigl(R+(\nabla\Phi)^2-8\Bigr).
\end{equation}
It is given by the following expressions for metric and dilaton field:
\begin{equation}\label{metric}
ds^2=\frac{dudv}{1-uv}, \qquad \phi=\log(1-uv).
\end{equation}

One should note that the metric has all the characteristic 
features of a black hole:
There is a horizon at $uv=0$ and a curvature singularity at $uv=1$. The 
following figure shows the Penrose diagram of the fully extended geometry
supplemented by regions behind the singularities (regions III and VI):
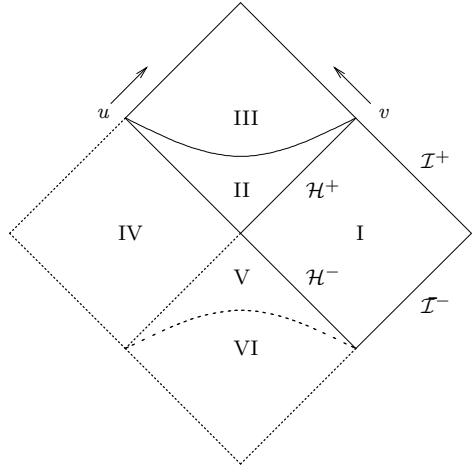
\begin{center}
\setlength{\unitlength}{0.0005in}
\begingroup\makeatletter\ifx\SetFigFont\undefined%
\gdef\SetFigFont#1#2#3#4#5{%
  \reset@font\fontsize{#1}{#2pt}%
  \fontfamily{#3}\fontseries{#4}\fontshape{#5}%
  \selectfont}%
\fi\endgroup%
{\renewcommand{\dashlinestretch}{30}
\begin{picture}(4824,4839)(0,-10)
\put(4362,1662){\makebox(0,0)[lb]{\smash{{{\SetFigFont{12}{14.4}{\rmdefault}{\mddefault}{\updefault}-}}}}}
\path(1062,3762)(1437,4137)
\path(1373.360,4030.934)(1437.000,4137.000)(1330.934,4073.360)
\path(3762,3762)(3387,4137)
\path(3493.066,4073.360)(3387.000,4137.000)(3450.640,4030.934)
\path(1212,3612)(2412,4812)(3612,3612)
	(4812,2412)(3612,1212)
\dottedline{45}(1212,3612)(12,2412)(1212,1212)
	(2412,12)(3612,1212)
\dottedline{45}(1212,1212)(2412,2412)
\path(2412,2412)(3612,3612)
\path(1212,3612)(1215,3611)(1221,3608)
	(1231,3603)(1247,3595)(1269,3584)
	(1296,3571)(1327,3556)(1362,3539)
	(1399,3521)(1436,3503)(1473,3486)
	(1510,3468)(1545,3452)(1578,3436)
	(1610,3422)(1639,3409)(1667,3396)
	(1693,3385)(1718,3374)(1742,3364)
	(1766,3355)(1789,3346)(1812,3337)
	(1835,3329)(1858,3320)(1882,3312)
	(1905,3304)(1930,3296)(1955,3289)
	(1980,3281)(2006,3274)(2032,3267)
	(2058,3260)(2085,3253)(2112,3247)
	(2139,3242)(2166,3237)(2192,3232)
	(2218,3228)(2244,3224)(2270,3221)
	(2294,3218)(2319,3216)(2342,3214)
	(2366,3213)(2389,3212)(2412,3212)
	(2435,3212)(2458,3213)(2482,3214)
	(2505,3216)(2530,3218)(2555,3221)
	(2580,3224)(2606,3228)(2632,3232)
	(2658,3237)(2685,3242)(2712,3247)
	(2739,3253)(2766,3260)(2792,3267)
	(2818,3274)(2844,3281)(2870,3289)
	(2894,3296)(2919,3304)(2942,3312)
	(2966,3320)(2989,3329)(3012,3337)
	(3035,3346)(3058,3355)(3082,3364)
	(3106,3374)(3131,3385)(3157,3396)
	(3185,3409)(3214,3422)(3246,3436)
	(3279,3452)(3314,3468)(3351,3486)
	(3388,3503)(3425,3521)(3462,3539)
	(3497,3556)(3528,3571)(3555,3584)
	(3577,3595)(3593,3603)(3603,3608)
	(3609,3611)(3612,3612)
\dottedline{450}(1212,1212)(1231,1221)(1296,1253)(1399,1303)(1510,1356)
(1610,1402)(1693,1439)(1766,1469)(1835,1495)(1905,1520)(1980,1543)(2058,1564)
(2139,1582)(2218,1596)(2294,1606)(2366,1611)(2435,1612)(2505,1608)(2580,1600)
(2658,1587)(2739,1571)(2818,1550)(2894,1528)(2966,1504)(3035,1478)(3106,1450)
(3185,1415)(3279,1372)(3388,1321)(3497,1268)(3577,1229)(3609,1213)
\put(3612,2337){\makebox(0,0)[lb]{\smash{{{\SetFigFont{8}{14.4}{\familydefault}{\mddefault}{\updefault}I}}}}}
\put(2337,3537){\makebox(0,0)[lb]{\smash{{{\SetFigFont{8}{14.4}{\familydefault}{\mddefault}{\updefault}III}}}}}
\put(1137,2337){\makebox(0,0)[lb]{\smash{{{\SetFigFont{8}{14.4}{\familydefault}{\mddefault}{\updefault}IV}}}}}
\put(2337,1137){\makebox(0,0)[lb]{\smash{{{\SetFigFont{8}{14.4}{\familydefault}{\mddefault}{\updefault}VI}}}}}
\put(2337,1887){\makebox(0,0)[lb]{\smash{{{\SetFigFont{8}{14.4}{\familydefault}{\mddefault}{\updefault}V}}}}}
\put(912,3612){\makebox(0,0)[lb]{\smash{{{\SetFigFont{8}{14.4}{\rmdefault}{\mddefault}{\itdefault}u}}}}}
\put(3837,3612){\makebox(0,0)[lb]{\smash{{{\SetFigFont{8}{14.4}{\rmdefault}{\mddefault}{\itdefault}v}}}}}
\put(2337,2787){\makebox(0,0)[lb]{\smash{{{\SetFigFont{8}{14.4}{\familydefault}{\mddefault}{\updefault}II}}}}}
\put(3087,1887){\makebox(0,0)[lb]{\smash{{{\SetFigFont{8}{14.4}{\rmdefault}{\mddefault}{\updefault}$\CH^-$}}}}}
\put(3087,2787){\makebox(0,0)[lb]{\smash{{{\SetFigFont{8}{14.4}{\rmdefault}{\mddefault}{\updefault}$\CH^+$}}}}}
\put(4287,3087){\makebox(0,0)[lb]{\smash{{{\SetFigFont{8}{14.4}{\rmdefault}{\mddefault}{\updefault}$\CI^+$}}}}}
\put(4287,1587){\makebox(0,0)[lb]{\smash{{{\SetFigFont{8}{14.4}{\rmdefault}{\mddefault}{\updefault}$\CI^-$}}}}}
\path(1212,3612)(3612,1212)
\end{picture}
}\end{center}
In the present paper only regions I, II and III will be considered, which is 
why the rest is represented by dotted lines. One may consider the metric
in regions I and II as an idealization of a black hole formed by gravitational
collapse if one imposes the boundary condition that there is no flux of
matter from regions IV and V into II and I respectively, cf. \cite{Un}.

The corresponding wave operator reads
\begin{equation}\label{waveop}
\begin{aligned}
\square=&-\frac{1}{2e^{\Phi}\sqrt{-g}}
\pa_{\mu}e^{\Phi}g^{\mu\nu}\sqrt{-g}\pa_{\nu} \\
 =& -\frac{1}{2}\bigl(\pa_u(1-uv)\pa_v+\pa_v(1-uv)\pa_u\bigr).
\end{aligned}\end{equation}

\subsection{Solutions to the wave equation in region I}
The wave equation to be solved reads
\begin{equation}\label{waveeq} 
\square f=\bigl(m^2-\fr{1}{4}\bigr)f.
\end{equation}
Splitting off $\frac{1}{4}$ on the right hand side is necessary for the mass
$m$ to be the mass as defined by an asymptotic observer.

In order to solve the wave-equation \rf{waveeq} it is useful to 
introduce variables $r=\log(-uv)$, $t=\log(-u/v)$ in which 
the covariant wave operator 
$\square$ takes the form
\begin{equation}\label{dalem} 
\square=e^{-r}\pa_r^{}(1+e^{r})\pa_r^{} - (e^{-r}+1)\pa_t^2.
\end{equation}
The operator $\De$ can furthermore be brought into the form of a 
one-dimensional Schr\"{o}dinger operator by considering the field $g=
e^{\Phi/2}f=\sqrt{1+e^r}f$ instead of $f$. The wave equation for
$g$ then reads
\[ \bigl(\pa_t^2+\De'\bigr)g=0,\qquad
\De'=-\pa_r^2 + V(r),\qquad V(r)=\frac{e^r}{4(1+e^r)^2}+m^2
\frac{e^r}{1+e^r}.
\]
Any solution which at some fixed time $t$ allows expansion of $g(r,t)$ and
$\dot{g}(r,t)\equiv \pa_t g(r,t)$ into generalized
eigenfunctions of the Schr\"{o}dinger operator $\De'$ 
can then be written in the form
\[ 
g(r,t)=\int d\mu(\om) \;\; \bigl(e^{-i\om t} W_{\om}(r)+
e^{+i\om t} \bar{W}_{\om}(r)\bigr) \qquad\text{with}\qquad
\De_h  W_{\om}(r) = \om^2 W_{\om}(r).
\]
The eigenvalue equation for $W_{\om}(r)$ is brought into
the form of a hypergeometric differential equation by $W_{\om}(r)=(-x)^{i\om}
(1-x)^{-1/2}F_{\om}(x)$, $x=-e^r$. 
One has two linearly independent solutions:
\[
U_{k}(r)= N_{k} e^{-i\om r}(1+e^r)^{\frac{1}{2}}
           F\bigl(\fr{1}{2}+i(k-\om),\fr{1}{2}-i(k+\om),1-2i\om,-e^r)
\]
and its complex conjugate $V_k(r)\equiv\bar{U}_k(r)$, 
where $k$ is fixed by the mass-shell condition 
$\om^2-k^2=m^2$, and $N_k$ is a normalization factor to be fixed below.
The asymptotic behavior for $r\ra \infty$ corresponding to large spacelike
distance from the black hole is given by plane waves:
\begin{equation}\label{asym} 
U_k(x) \sim N_k
\bigl(B_{+}(k)e^{-ikr}+B_{-}(k)e^{ikr}\bigr),\qquad
B_{\pm}(k)=\frac{\Ga(1-2i\om)\Ga(\mp 2ik)}{\Ga^2\bigl(\frac{1}{2}-i(\om
\pm k)\bigr)}.
\end{equation}
It can then be checked that $\De'$ is essentially 
self-adjoint in $L^2(\BR)$:
There are no square-integrable eigenfunctions of $\De'$, so the deficiency 
indices vanish. It can furthermore be shown that
the set of generalized eigenfunctions 
$\{U_k(x); k\in\BR_+\}\cup\{V_k(x); k\in\BR_+\}$ constitutes
a plane wave basis for $L^2(\BR)$. The normalization $N_k$ is finally
given by
\[
\int_{\BR}dr \;\;\bar{U}_{k_2}(r) U_{k_1}(r)=2\pi\de(k_2-k_1),
\qquad
\int_{\BR}dr \;\;\bar{U}_{k_2}(r) V_{k_1}(r) = 0,
\]
if the normalization $N_{k}$ is chosen as
\[
N_{k}=\frac{1}{B_{+}(k)}=\frac{\Ga^2
\bigl(\frac{1}{2}-i(k+\om)\bigr)}{\Ga(1-2i\om)\Ga(-2ik)},
\]
corresponding to normalizing the ``incoming'' plane wave in \rf{asym}
to unity.

This yields existence and uniqueness
of a solution to the Cauchy-problem for suitable
subspaces $\CS \subset L^2(\BR)$ of test-functions, where $\CS$ could be
for example the usual Schwartz space of $L^2(\BR)$. It takes the form
\begin{equation}\label{exp}
g(r,t)=
\int_0^{\infty}dk\Bigl(   e^{-i\om t}\bigl(
a_kU_{k}(r)+b_k V_{k}(r)\bigr)  
+   e^{i\om t}\bigl(
\ba_k\bar{U}_{k}(r)+\bb_k \bar{V}_{k}(r)\bigr)\Bigr),
\end{equation}
where the coefficients are given
in terms of the values $g(r,t)$, $\dot{g}(r,t)$ at fixed
time $t$ via 
\begin{equation}\label{inifourier}
\begin{aligned}
a_k=a_k[g,\dot{g}]=& 
\frac{1}{4\pi}\int_{\BR}dr \;\;e^{i\om t}\bar{U}_{k}(r)\Bigl(g(r,t)+
\frac{i}{\om}\dot{g}(r,t)\Bigr) \\[1ex]
b_k=b_k[g,\dot{g}]=& 
\frac{1}{4\pi}\int_{\BR}dr \;\;e^{i\om t}\bar{V}_{k}(r)\Bigl(g(r,t)-
\frac{i}{\om}\dot{g}(r,t)\Bigr). 
\end{aligned} 
\end{equation}

In view of a similar phenomenon that will be found in the deformed case
it may be worthwhile noting that expansion into eigenfunction of $\De'$
is possible for considerably more general choices of 
subspaces $\CT$ , $\CS\subset \CT \subset L^2(\BR)$
\footnote{see the first two 
sections of \cite{Be} for a lucid discussion of the conditions $\CT$ has to
satisfy in order to allow expansion of {\it any} $g\in\CT$ into 
generalized eigenfunctions}. Moreover, the expression \rf{exp}
still makes sense for $a_k=a_k[g,\dot{g}]$, $b_k=b_k[g,\dot{g}]$ corresponding
to initial data $g$, $\dot{g}$ in $\CT$. But the resulting 
function $g(r,t)$ will then generically not be differentiable; it can 
therefore be considered as a solution of the wave-equation only in 
the distributional sense. This 
phenomenon is familiar from the simple case $\pa_t^2f=\pa_x^2f$: 
One may  consider $f(x-t)$ to be a distributional solution of 
$\pa_t^2f=\pa_x^2f$ even if $f$ is not differentiable.

\subsection{Scattering in the black hole geometry}

By the method of stationary phase it is possible to 
show that for  $t\ra -\infty$ 
any solution \rf{exp} is asymptotic to a function $\gas(r,t)$
in the sense that
\[ 
\lim_{t\ra -\infty}\int_{\BR}dr\;\;\lvert  g(r,t)-\gas(r,t)\rvert^2 =0.
\]
The function $\gas(r,t)$ is expressed in terms of $a_k$, $b_k$ as follows
\[
\gas=\gas_1+\gas_2\qquad\begin{aligned}
 \gas_1(r,t)=&\int_{0}^{\infty}dk \;\bigl(e^{-i\om(t-r)}b_k \bar{N}_k  
+ e^{i\om(t-r)}\bb_k N_k\bigr)\\[1ex]
 \gas_2(r,t)=&\int_{0}^{\infty}dk \;\bigl(e^{-i(\om t+k r)}
(a_k +b_k)+ e^{i(\om t+k r)}
(a_k +b_k)\bigr).
\end{aligned}\]
The functions $\gas_1$ ($\gas_2$) 
describe right- (left-) moving wave-packets
coming in from $\CH_{-}$ ($\CI_{-}$). 

However, the right-moving 
plane waves at $\CH_{-}$ represent an inflow from region $V$
into region $I$. In order to be consistent with the interpretation of
regions I/II as being an idealization of a black hole formed by 
gravitational collapse, it is necessary to impose the boundary condition
of vanishing $\gas_1$ corresponding to $b_k\equiv 0$.

The scattering problem for a wave-packet with asymptotic form $\gin$ 
consists therefore in determining the late-time
asymptotics $\gout(r,t)$ defined by
\[
\lim_{t\ra \infty} \int_{\BR}dr\;\;
\lvert  g(r,t)-\gout(r,t)\rvert^2 =0.
\]
for $f(r,t)$ subject to the boundary condition $b_k=0$. It is given by
$\gout(r,t)=\gout_1(r,t)+\gout_2(r,t)$:
\[ \begin{aligned}
\gout_1(r,t)=& \int_{m}^{\infty}d\om\;\; 
\bigl(e^{-i\om(t+r)}T_ka_k + e^{i\om(t+r)}\bar{T}_k\ba_k \bigr)\\[1ex]
\gout_2(r,t)=&\int_{0}^{\infty}dk 
\;\bigl( e^{-i(\om t+k r)}R_ka_k+e^{i(\om t+k r)}\bar{R}_k\ba_k\bigr),
\end{aligned}
\]
where $\gout_1(r,t)$ describes the matter that falls through the future 
horizon $\CH_+$,
whereas $\gout_2(r,t)$ represents the part that escapes towards 
space-like infinity. The corresponding
``transmission'' amplitude $T_k$ and ``reflection amplitude $R_k$ are 
respectively given by
\[
T_k= \frac{\om}{k}N_k= \frac{\Ga^2
\bigl(\frac{1}{2}-i(k+\om)\bigr)}{\Ga(-2i\om)\Ga(1-2ik)}\qquad
R_k= \frac{N_k}{N_{-k}}=\frac{\Ga^2
\bigl(\frac{1}{2}-i(k+\om)\bigr)}
{\Ga^2\bigl(\frac{1}{2}+i(k-\om)\bigr)}
\frac{\Ga(+2ik)}{\Ga(-2ik)}
\]
 Note that information is conserved: $|T_k|^2+|R_k|^2=1$.

\subsection{Continuation into region II}
The continuation of the wave-packet \rf{exp} into region
II is trivial when expressing the modes $e^{-i\om t}U_{k}(r)$ 
in terms of $u$, $v$-coordinates:
\begin{equation}\label{contII}
e^{-i\om t}U_{k}(r)\equiv \CU_{k}(u,v)=u^{-2i\om}F\bigl(
\fr{1}{2}+i(k-\om),\fr{1}{2}-i(k+\om),1-2i\om,uv\bigr),
\end{equation}
where $\CU_{k}(u,v)$ is analytic in $v$ near the horizon $v=0$. The expression
\[
f(u,v)=\int_m^{\infty}d\om \;\;\bigl(\CU_{k}(u,v)T_{k}a_k +
\bar{\CU}_{k}(u,v)\bar{T}_{k}\ba_k\bigr)
\]
therefore defines a wave-packet $f$ in the union of regions I and II. This 
continuation becomes unique by imposing the condition of vanishing 
on the boundary between regions IV and II, which is motivated by arguments
analogous to those that motivated vanishing on $\CH_-$.

However, the singularity at $uv=1$ prevents further continuation into region
III. Technically this follows from the fact that the modes
$\CU_{k}(u,v)$ develop a singularity of the form $\log(1-uv)$. 
Predictability of the evolution of the wave-packet $f$ breaks down at
the singularity
since there is no unique way of defining $\log(x)$ for negative $x$. 

\section{Propagation and scattering in the deformed black hole}

The deformation of the covariant wave-equation that will be considered
takes the form
\begin{equation}\label{q-box}
\square_h f=\bigl\{ m^2-\fr{1}{4}\bigr\}_h^{} f, \qquad 
\bigl\{ m^2-\fr{1}{4}\bigr\}_h^{}\equiv 
\biggl(\frac{\sinh^2(\pi h m)}{ \sin^2(\pi h)}
-\frac{\sin^2(\frac{\pi h}{2})}{\sin^2(\pi h)}
\biggr) ,
\end{equation}
where $h\in(0,1)$ is the deformation parameter and
the differential operator $\square$ that appeared in the classical
case \rf{dalem} has been replaced by the following {\it finite
difference} operator:
\[ \square_{h}=
e^{-r}D_r^{}(1+e^{r})D_r^{} - (e^{-r}+1)D_t^2\qquad
\begin{aligned}
D_r \equiv & 
\frac{\de_r^{+}-\de_r^{-}}{2i\sin(\pi h)},\qquad 
\de_r^{\pm}f(r,t)=f(r\pm\pi i h,t),\\
D_t \equiv &
\frac{\de_t^{+}-\de_t^{-}}{2i\sin(\pi h)},
\qquad \de_t^{\pm}f(r,t)=f(r,t\pm\pi i h).
\end{aligned}\]
One obviously recovers the classical
counterparts in the limit $h\ra 0$. What appears to be unusual about this
kind of deformation is the appearance of {\it imaginary} shifts of the
arguments $r,t$. In order for operators such as $\square_{h}$ to be 
defined as operators on spaces of functions of {\it real}
variables $r,t$ one needs an unambigous prescription to extend the
functions defined for real values of the arguments to the relevant strips
in the complex plane. The most natural such extension seems to be given
by requiring analyticity in the strip
\[ 
\CS=\bigl\{ (r,t)\in\BC^2 ; |\Im(r)|<2\pi h, \;\; |\Im(t)|<2\pi h \bigr\},
\]
and existence of the limits  
\[
f(r\pm 2\pi ih,t)=\lim_{\ep\ra 0} f(r\pm 2\pi ih\mp i\ep,t), \qquad 
f(r,t\pm 2\pi ih)=\lim_{\ep\ra 0} f(r,t\pm 2\pi ih\mp i\ep), \qquad \ep>0
\]
for almost any $r,t\in\BR$.
This choice can alternatively be justified by considering the reduction from
the quantized space $SL_q(2,\BR)$.

\subsection{Preliminaries}
It is useful to write \rf{q-box} in the form 
\[ 
(D_t^2+\De_h)f=0,\qquad  \De_h=-\frac{1}{1+e^r}D_r(1+e^r)D_r + 
\bigl\{ m^2-\fr{1}{4}\bigr\}_h^{}
\frac{e^r}{1+e^r},
\]
or in terms of $g=(1+e^r)^{\frac{1}{2}}f$
\[
(D_t^2+\De'_h)g=0,\qquad  \De'_h=-\frac{1}{\sqrt{(1+e^r)}}D_r(1+e^r)
D_r\frac{1}{\sqrt{(1+e^r)}} + \bigl\{ m^2-\fr{1}{4}\bigr\}_h^{}
\frac{e^r}{1+e^r}.
\]
As in the classical case one may try to construct a representation for the
general solution by means of an eigenfunction expansion for 
$\De_h$:\footnote{In this case it is technically more convenient to 
consider $\De_h$ instead of $\De_h'$} If
\[ 
f(r,t)=\int d\mu(\om) a_{\om}(t) W_{\om}(r) \qquad\text{with}\qquad
\De_h  W_{\om}(r) = [\om]_h^2 W_{\om}(r), \qquad
[z]_h\equiv \frac{\sinh(\pi h z)}{\sin(\pi h)} ,
\]
then $a_{\om}(t)$ will be determined as a solution of 
$D_t^2a_{\om}(t)=-[\om]_h^2a_{\om}(t) $. The most general
solution of the latter equation that has the required analyticity 
is given by 
\[
 a_{\om}(t) = \sum_{m\in\BZ} 
\bigl(A_m e^{-i(\om+i\ka m)t} + B_m e^{i(\om+i\ka m)t}\bigr).
\]
Note that the contributions from $m\neq 0$ blow up for $t\ra\infty$
or $t\ra -\infty$. They correspond to the ``runaway''-solutions that
typically cause problems for higher-derivative equations of motion. However, 
in the present case one has not any problem to simply throw away the
``badly-behaved'' solutions with $n\neq 0$. Together with the 
eigenfunction decomposition of $\De_h$ one will thereby obtain a perfectly
well-defined initial value problem:

\subsection{Eigenfunction expansion for $\De_h$}
$\De_h$ will be considered as an operator defined on the 
dense domain $\CD\subset L^2(\BR,d\la(r))$, $d\la(r)=dr(1+e^r)$
which consists of functions
that allow extension to a function 
holomorphic in the strip $\{ r\in\BC; |\Im(r)|< 2\pi h\}$
and satisfy
\[ 
\sup_{|\eta|<2\pi h}\int_{\BR}d\la(r)\;\;|f(r+i\eta)|^2
<\infty.
\] 
\begin{thm}
The operator $\De_h$ is essentially self-adjoint.
There exists an expansion into generalized eigenfunctions of $\De_h$.
\end{thm}

\begin{thm}
The following set $\FB$ constitutes a basis of generalized eigenfunctions
for $\De_h$:
\[ 
\FB=\bigl\{      U_{h,k}; k\in \BR_+\bigr\}\cup \bigl
         \{      V_{h,k}; k\in \BR_+\bigr\},
\]
where the eigenvalue $\om^2$ is given in terms of the parameter $k$ 
by the {\it h-mass-shell} relation
\begin{equation}\label{h-mass}
[\om]_h^2-[\ka]_h^2=[m]_h^2
\end{equation}
and $U_{h,k}(r)$ is given in terms of the
q-hypergeometric functions introduced in the Appendix (eqn. \rf{qhyp}) as
\[
U_{h,k}(r)=N_{h,k}^+ e^{-i\om r}
        F_h\bigl(\fr{1}{2}+i(k-\om),\fr{1}{2}-i(k+\om),1-2i\om,-e^r\bigr),
\]
whereas $V_{h,k}(r)=\bar{U}_{h,k}(r)$,the complex conjugate of $U_{h,k}(r)$. 
\end{thm}
 
\begin{thm} The functions $U_{h,k}(r)$, $\bar{U}_{h,k}(r)$ are orthonormalized
according to
\[
\int_{\BR}d\la(r)\;\;  \bar{U}_{h,k_2}(r) U_{h,k_1}(r)=2\pi\de(k_2-k_1),
\qquad
\int_{\BR}d\la(r)\;\;  \bar{V}_{h,k_2}(r) U_{h,k_1}(r) = 0,
\]
if the normalization $N_{h,k}$ is chosen as
\[
N_{h,k}=\frac{\Ga_h^2
\bigl(\frac{1}{2}-i(k+\om)\bigr)}{\Ga_h(1-2i\om)\Ga_h(-2ik)}.
\]
\end{thm}
The normalization is as in the classical case such that the ``incoming wave''
in the ($r\ra\infty$)-asymptotics is normalized to one:
\[ 
U_{h,k}(r)\sim e^{-\frac{1}{2}r}\biggl(e^{-ikr} + \frac{N_{h,k}}{N_{h,-k}}
e^{+ikr}\biggr).
\]

Although these theorems are precise analogues
of the corresponding statements in the undeformed case, their proof is quite
different. To the authors knowledge these are the first nontrivial 
results on spectral analysis of finite difference operators of this type.

\subsection{Initial value problem}
The results just given allow one to write any solution as
\begin{equation}\label{q-exp}
f(r,t)=
\int_0^{\infty}dk\Bigl(   e^{-i\om t}\bigl(
a_k U_{h,k}(r)+b_k V_{h,k}\bigr)  
+   e^{i\om t}\bigl(
\ba_k\bar{U}_{h,k}(r)+\bb_k \bar{V}_{h,k}\bigr)\Bigr),
\end{equation}
where the coefficients are given
in terms of the values $f(r,t)$, $\dot{f}(r,t)\equiv\pa_t f(r,t)$ at fixed
time $t$ via 
\begin{equation}\label{q-inifourier}
\begin{aligned}
a_k=a_k[f,\dot{f}]=& 
\frac{1}{4\pi}\int_{\BR}d\la(r) \;\;e^{i\om t}\bar{U}_{h,k}(r)\Bigl(f(r,t)+
\frac{i}{\om}\pa_t f(r,t)\Bigr), \\[1ex]
b_k=b_k[f,\dot{f}]=& 
\frac{1}{4\pi}\int_{\BR}d\la(r) \;\;e^{i\om t}\bar{V}_{h,k}(r)\Bigl(f(r,t)-
\frac{i}{\om}\pa_t f(r,t)\Bigr). 
\end{aligned} 
\end{equation}
However, it can be shown that the expression \rf{q-exp} remains sensible
for much more general choices of the coefficients $a_k$, $b_k$ 
than those provided by \rf{q-inifourier} for solutions $f(r,t)$.
Stated differently, 
general choice of $a_k$, $b_k$ in \rf{q-exp} will not yield $f$ 
that have the analyticity properties necessary to be solutions in the
strict sense, but only in the distributional sense.\footnote{What 
appears to be at work is a generalization of the Paley-Wiener
theorems relating analyticity of functions on a strip to exponential decay
properties of their Fourier transforms.}
  
On the other hand one may observe that the function $f(r,t)$ given by 
\rf{q-exp} can alternatively be 
characterized as a solution of a {\it second order
differential} equation w.r.t. time of the form
\begin{equation}\label{q-evol}
(\pa_t^2+\CD_h)f=0, \qquad \bigl(\CD_h f\bigr)(r,t)=
\int_{\BR}dr' \;\CK_h(r,r')f(r',t)
\end{equation} 
where the kernel $\CK_h(r,r')$ is given by 
\begin{equation}\label{dkernel}
d_h(r,r')=\frac{1}{2\pi}\int_0^{\infty}dk \;\;\om^2\;\;
\bigl(U_{h,k}(r)\bar{U}_{h,k}(r)+ V_{h,k}(r)\bar{V}_{h,k}(r)\bigr),
\end{equation}
and $\om$ has to be expressed in terms of $k$ by means of the h-mass-shell
relation. The corresponding expression $d(r,r')$ in 
the classical case is of course simply equal to $(-\pa_r^2+V(r))\de(r-r')$. 
The fact that $d(r,r')$ is supported on the diagonal can e.g. be found by 
extending the $k$-integration in the classical analogue of \rf{dkernel}
to the full axis and closing the contour in the upper (lower) half-plane 
depending on the sign of $r-r'$. This will no longer be possible in the 
deformed case since $\om$ as function of $k$
has logarithmic and square-root branch cuts. $\CD_h$ is therefore most likely
a genuinely nonlocal operator.

To summarize: The deformed wave equation supplemented
with the condition of absence of ``runaway''-solutions is equivalent to
the nonlocal evolution equation \rf{q-evol}, which manifestly has a
well-posed initial value problem.

\subsection{Scattering in region I}
At this point it becomes possible to carry through a discussion of scattering
in region I of the deformed black hole in complete analogy to the undeformed
case. It basically amounts to adding subscript ``h'' at the appropriate 
places. 

First of all it turns out that the boundary condition of vanishing on the
past horizon $\CH_-$ again corresponds to $b_k\equiv 0$ in \rf{q-evol}. Then
one finds 
\begin{thm} The asymptotics $\gin$, $\gout$ for $t\ra \mp\infty$ defined by 
\[
\lim_{t\ra\mp\infty}\int_{\BR}dr |g(r,t)-g^{\text{\tiny in/out}}
(r,t)|^2 =0
\] 
is explicitely given by
\[ \begin{aligned} 
\gin(r,t)=& \int_0^{\infty} dk\;\; \bigl( e^{-i(\om t+kr)}a_k+
e^{i(\om t+kr)}\ba_k \bigr),\\
\gout(r,t)=& \gout_1(r,t)+ \gout_2(r,t),\qquad 
\begin{aligned}
\gout_1(r,t)=& \int_m^{\infty}d\om \bigl(e^{-i\om(t+r)}T_{h,k}a_k+
e^{+i\om(t+r)}\bar{T}_{h,k}\ba_k\bigr), \\[1ex]
\gout_2(r,t)=& \int_0^{\infty}dk   \bigl(e^{-i(\om t+k r)}R_{h,k}a_k+
e^{+i(\om t-kr)}\bar{R}_{h,k}\ba_k\bigr),
\end{aligned}\end{aligned}
\]
where the reflection and transmission coefficients are given
by
\[ \begin{aligned}
T_{h,k}=&\frac{{[}2\om{]}_h}{{[}2k{]}_h}N_{h,k}=\frac{\Ga_h^2
\bigl(\frac{1}{2}-i(k+\om)\bigr)}{\Ga_h(-2i\om)\Ga_h(1-2ik)}\\
R_{h,k}=&\frac{N_{h,k}}{N_{h,-k}}=\frac{\Ga_h^2
\bigl(\frac{1}{2}-i(k+\om)\bigr)}
{\Ga_h^2\bigl(\frac{1}{2}+i(k-\om)\bigr)}
\frac{\Ga_h(+2ik)}{\Ga_h(-2ik)}
\end{aligned}
\]
\end{thm}
It is noteworthy that one has {\it ordinary plane waves} in the
asymptotic regions! This can be understood by noting that
for scales in $r$, $t$-space that are
large compared to $h$ and $\om$, $k$, $m$ small compared to $h$ one
may approximate the deformed wave equation by the undeformed one. 
The asymptotic observer will see the effect of deformation only by 
analyzing high frequencies of the reflected waves. 

Finally, one may again check that information is preserved: 
$|T_{h,k}|^2+|R_{h,k}|^2=1$. 

\subsection{Continuation into regions II/III}
A remarkable qualitative difference to the undeformed case shows up 
in considering the continuation of the wave-packet that passes through the
horizon into regions II/III. To this aim one should again use the $u$, $v$-
coordinates. In terms of these one has 
\begin{equation}\label{q-contII}
e^{-i\om t}U_{h,k}(r)\equiv \CU_{h,k}(u,v)=u^{-2i\om}F_h\bigl(
\fr{1}{2}+i(\ka-\om),\fr{1}{2}-i(\ka+\om),1-2i\om,uv\bigr),
\end{equation}
which is continuously differentiable w.r.t. $v$ on the future horizon 
$v=0$.\footnote{Cf. Proposition 2.1. of the Appendix. Here it is important
to restrict $h$ to be in $(0,1)$} Wave packets
of these modes therefore have a well-defined continuation into region II.
As in the classical case one gets a unique solution in region II 
by demanding vanishing on the boundary between regions II and IV. Explicitly
it reads
\begin{equation}\label{wavpkII}
f_{II}(u,v)=\int_0^{\infty}dk \;\;\bigl(\CU_{h,k}(u,v)T_{h,k}a_k +
\bar{\CU}_{h,k}(u,v)\bar{T}_{h,k}\ba_k\bigr).
\end{equation}

But what appears to be remarkable is the fact that the modes $\CU_{h,k}(u,v)$
are nonsingular for any $u,v>0$: The singularity has disappeared. In fact,
the integral \rf{qhyp} that defines the q-hypergeometric function in 
\rf{q-contII}
converges absolutely for any positive as well as negative values of $uv$. 
Using the variable $\rho=\log(uv)$ in region II/III one finds that the
singularity that classically was at $uv=1$ resp. $\rho=0$ now has been
resolved into a series of poles at $\rho=i(nh+(m-1))$, $n,m=1,2,\ldots$.
These poles approach the real axis in the classical limit $h\ra 0$ .

So what is the fate of matter fallen into the black hole in the deformed case?
The further propagation of \rf{wavpkII} in regions II/III
can be described in terms of the time variable
$\tau=\log(u/v)$ the
same way as was discussed in region I. The late time asymptotics
of a wave-packet that has fallen into the black hole is then given by 
\[ \begin{aligned}
\fout_{II}(\rho,\tau)= & \int_0^{\infty}dk\;\;\biggl(
\Bigl(e^{-i(\tau \om+ k\rho)} S_{h,k}^+ + e^{+i(\tau \om+ k\rho)}
\bar{S}_{h,k}^+\Bigr)+\Bigl(
e^{-i(\tau \om- k\rho)} S_{h,k}^- + e^{+i(\tau \om- k\rho)}\bar{S}_{h,k}^- 
\Bigr)\biggr),\\
 & \text{where}\quad 
S_{h,k}^+=e^{-\frac{\pi i}{2}}e^{-\pi(\om-k)}\qquad\qquad
S_{h,k}^-=e^{-\frac{\pi i}{2}}e^{-\pi(\om+k)}R_{h,k}.
 \end{aligned} \]

\section{Conclusions}

In the author's opinion there are three main lessons to learn from the
example studied in the present paper:
\begin{enumerate}
\item There are nonlocal evolution laws that may be interpreted as 
describing propagation of fields on deformed geometries which 
allow one to avoid some of the usual problems 
associated with nonlocalities in a natural way.
\item This way of deformation indeed provides a resolution of singularities
present in the classical geometry, which allows one to
propagate wave-packets through the region that classically was 
singular. Such deformations may therefore 
open ways to resolve the black hole information problem.
\item There are ways to describe some kinds of
small-scale ``fuzziness'' or nonlocality that do not require 
non-commutativity of the underlying algebra of functions.
\end{enumerate}

The example presented here is of course somewhat artificial in being 
distinguished by its simplicity and solvability. Its main value is to 
illustrate the above-mentioned qualitative points which one may expect
to persist in considerably more general types of deformations.

More can be done along similar lines as presented here: First 
one may observe that the present discussion already contains important
ingredients for studying the quantization of solutions of the deformed
wave equation, with the aim of ultimately determining how the
effect of deformation shows up in the spectrum of the Hawking-radiation.

Furthermore, it was mentioned in the introduction that the present
model is just one case of a class of models that can be constructed and 
investigated along similar lines. In contrast to the present one however,
the other models are all noncommutative deformations. 

Of particular interest may be to study the deformation of $SL(2,\BR)\simeq
ADS_3$ as a model for the possible nonlocality (e.g. \cite{BDHM})
of string theory on backgrounds with $AdS_3$, similarly to what was 
recently proposed in \cite{JR}.

Finally it should be emphazized that the real task remains  
to find more concrete evidence on the small scale structure of space-time
from the study of the full-fledged gravitational theories such as 
string theory.
 
\section{Appendix: $q$-special functions for $q=e^{\pi i h}$}
\subsection{q-Gamma function}
The basic building block for the class of special functions to be considered
is the the Double Gamma function introduced by Barnes \cite{Ba}
\begin{defn} The Double Gamma function is defined as
\[
\log\Ga_2(s|\om_1,\om_2)=  \Biggl(\frac{\pa}{\pa t}\sum_{n_1,n_2=0}^{\infty}
(s+n_1\om_1+n_2\om_2)^{-t}\Biggr)_{t=0}.
\]
\end{defn}
\begin{defn}
The h-Gammafunction $\Ga_h$:
\[
\quad \Ga_h(s)=\frac{\Ga_2\bigl(s|1,\ka)}
{\Ga_2\bigl(1+\ka-s|1,\ka\bigr)},\qquad \ka=h^{-1}, 
\] 
\end{defn}
\begin{prop} Properties:
\begin{enumerate}
\item Functional relations:
\[ \Ga_h(s+1)=2\sin(\pi hs)\Ga_h(s) \qquad \Ga_h(s+\ka)=
2\sin(\pi s)\Ga_h(s)
\]
\item Zeros at $s=1+\ka+n+m\ka$, Poles at $s=s_{n,m}=-n-m\ka$, 
$n,m=0,1,2,\ldots $.
\[
\lim_{s\ra s_{mn}}s \Ga_h(s)=\frac{1}{2\pi b}
\frac{(-)^{n+m+mn}}{[n]!_h[m]!_{h^{-1}}} \qquad [n]_h!=
\prod_{r=1}^{n}(q^r-q^{-r})
\]
\item Asymptotics: For $|s|\ra\infty$ in any sector not containing the
real line one has
\[
\log \Ga_h(s)\sim \mp 
\frac{\pi ih}{2}\bigl(s^2-s(1+\ka)\bigr) +\CO(s^{-1})
\quad\text{for}\quad \pm\Im(s)>0 \]
\end{enumerate}\end{prop}
Proof: \cite{Sh}

\begin{defn} The q-hypergeometric function will be defined as
\begin{equation}\label{qhyp}
F_h(\al,\be;\ga;z)=\frac{\Ga_h(\ga)}{\Ga_h(\al)\Ga_h(\be)}
\int_{-i\infty}^{i\infty}ds 
\frac{(-z)^s}{\sin(\pi s)} 
\frac{\Ga_h(\al+s)\Ga_h(\be+s)}{\Ga_h(\ga+s)
\Ga_h(1+s) },
\end{equation}
where the contour is to the right of the poles at $s=-\al-n-m\ka$
$s=-\be-n-m\ka$ and to the left of the poles at $s=n+m\ka$
$s=1+\ka-\ga+n+m\ka$, $n,m=0,1,2,\ldots$.
\end{defn}
This definition of a q-hypergeometric function is closely related to
the one first given in \cite{NU}.
\begin{prop} Properties:
\begin{enumerate}
\item Asymptotic behavior for $x\ra 0$
\[ 
F_h(\al,\be,\ga;z)=  
1+\CO(z)+\frac{\Ga_h(1+\ka+\al-\ga)\Ga_h(1+\ka+\al-\ga)
\Ga_h(\ga)}{\Ga_h(\al)\Ga_h(\be)\Ga_h(2+2\ka-\ga)}
(-z)^{1+\ka-\ga} 
\bigl(1+\CO(z)\bigr)
\]
\item Asymptotic behavior for $x\ra-\infty$
\[
F_h(\al,\be,\ga;z)=
\frac{\Ga_h(\ga)\Ga_h(\be-\al)}{\Ga_h(\be)\Ga_h(\ga-\al)}
(-z)^{-\al}\bigl(1+\CO(z^{-1})\bigr)+
 \frac{\Ga_h(\ga)\Ga_h(\al-\be)}{\Ga_h(\al)\Ga_h(\ga-\be)}
(-z)^{-\be}\bigl(1+\CO(z^{-1})\bigr)
\]
\end{enumerate}\end{prop}
\newcommand{\CMP}[3]{{\it Comm. Math. Phys. }{\bf #1} (#2) #3}
\newcommand{\LMP}[3]{{\it Lett. Math. Phys. }{\bf #1} (#2) #3}
\newcommand{\IMP}[3]{{\it Int. J. Mod. Phys. }{\bf A#1} (#2) #3}
\newcommand{\NP}[3]{{\it Nucl. Phys. }{\bf B#1} (#2) #3}
\newcommand{\PL}[3]{{\it Phys. Lett. }{\bf B#1} (#2) #3}
\newcommand{\MPL}[3]{{\it Mod. Phys. Lett. }{\bf A#1} (#2) #3}
\newcommand{\PRL}[3]{{\it Phys. Rev. Lett. }{\bf #1} (#2) #3}
\newcommand{\AP}[3]{{\it Ann. Phys. (N.Y.) }{\bf #1} (#2) #3}
\newcommand{\LMJ}[3]{{\it Leningrad Math. J. }{\bf #1} (#2) #3}
\newcommand{\FAA}[3]{{\it Funct. Anal. Appl. }{\bf #1} (#2) #3}
\newcommand{\PTPS}[3]{{\it Progr. Theor. Phys. Suppl. }{\bf #1} (#2) #3}
\newcommand{\LMN}[3]{{\it Lecture Notes in Mathematics }{\bf #1} (#2) #2}


\begin{thebibliography}{99}
\bibitem[DFR]{DFR}
S. Doplicher, K. Fredenhagen, J.E. Roberts:
The quantum structure of spacetime of the Planck scale and quantum fields.
Commun. Math. Phys. {\bf 172}(1995) 187-220 
\bibitem[FGR]{FGR} J. Fr\"{o}hlich, O. Grandjean, A. Recknagel: 
   Supersymmetric quantum theory, non-commutative geometry, and gravitation,
   in: Proceedings of the Les Houches Summer School 1995, Session LXIV, 
   North-Holland, Elsevier (1998)
\bibitem[DKPS]{DKPS}  M.R. Douglas, D. Kabat, P. Pouliot, S.H. Shenker:
  D-branes and Short Distances in String Theory, Nucl.Phys.B 
  {\bf 485}(1997) 85-127
\bibitem[D]{D} M.R. Douglas: Two Lectures on D-Geometry and Noncommutative 
Geometry, hep-th/9901146
\bibitem[BDHM]{BDHM}  T. Banks, M.R. Douglas, G.T. Horowitz, E. Martinec:
  AdS Dynamics from Conformal Field Theory, hep-th/9808016
\bibitem[Co]{Co} A. Connes: Noncommutative geometry. 
San Diego, CA: Academic Press (1994)
\bibitem[FG]{FG} J. Froehlich, K. Gawedzki:
  Conformal field theory and geometry of strings. In:
  Feldman, J. (ed.) et al., Mathematical quantum theory I: 
  Field theory and many body theory. Proceedings of the Canadian 
  Mathematical Society annual seminar, Vancouver, August 4-14, 1993. 
  Providence, RI: American Mathematical Society, CRM Proc. Lect. Notes. 7,
  (1994) 57-97 
\bibitem[Wi]{Wi} E. Witten: On string theory and black holes,
Phys.Rev.D{\bf 44} (1991) 314-324. 
\bibitem[DVV]{DVV} R. Dijkgraaf, H. and E. Verlinde: String 
  propagation in a black hole geometry, \NP{371}{1992}{269-314}
\bibitem[Un]{Un} W. Unruh: Notes on black hole evaporation,
Phys.Rev.D{\bf 14} (1976) 870
\bibitem[Be]{Be} J. Bernstein: On the support of Plancherel measure,
J. Geom. Phys. {\bf 5} (1988) 663-710 .
\bibitem[JR]{JR} A. Jevicki, S. Ramgoolam: Non-commutative gravity from the 
ADS-CFT correspondence, hep-th/9902059
\bibitem[Ba]{Ba} E.W. Barnes: Theory of the double gamma function, Phil. 
Trans. Roy. Soc. A {\bf 196} (1901) 265-388
\bibitem[Sh]{Sh} T. Shintani: On a Kronecker limit formula for real
quadratic fields, J. Fac. Sci. Univ. Tokyo Sect.1A {\bf 24}(1977)167-199
\bibitem[NU]{NU} M. Nishizawa, K. Ueno: Integral soluitons of q-difference 
equations of the hypergeometric type with $|q|=1$,  q-alg/9612014  
\end{thebibliography}
\end{document}